\UseRawInputEncoding 
\documentclass[twocolumn]{aastex631}

\usepackage[mathscr]{euscript}
\usepackage{bm}

\mathchardef\mhyphen="2D

\newcommand{\mearth}{~\rm{M}_{\oplus}}
\newcommand{\rearth}{\rm{R}_{\oplus}}
\newcommand{\mjup}{~\rm{M}_{\rm Jup}}

\newcommand{\eforced}{e_{\rm{forced}}^{\rm{cd}}}
\newcommand{\efree}{e_{\rm{free}}^{\rm{cd}}}

\shorttitle{V1298 Tau stability constraints}
\shortauthors{Tejada Arevalo et al.}
\graphicspath{{./}{figures/}}

\begin{document}

\title{Stability Constrained Characterization of the 23 Myr-old V1298 Tau System: Do Young Planets Form in Mean Motion Resonance Chains?}

\author[0000-0001-6708-3427]{Roberto Tejada Arevalo}
\affiliation{Department of Astrophysical Sciences, Princeton University, 4 Ivy Ln, Princeton, NJ 08544, USA}

\author[0000-0002-9908-8705]{Daniel Tamayo}
\affiliation{Department of Astrophysical Sciences, Princeton University, 4 Ivy Ln, Princeton, NJ 08544, USA}

\author[0000-0002-6458-3423]{Miles Cranmer}
\affiliation{Department of Astrophysical Sciences, Princeton University, 4 Ivy Ln, Princeton, NJ 08544, USA}

\begin{abstract}
A leading theoretical expectation for the final stages of planet formation is that disk migration should naturally drive orbits into chains of mean motion resonances (MMRs). 
In order to explain the dearth of MMR chains observed at Gyr ages ($<1\%$), this picture requires such configurations to destabilize and scramble period ratios following disk dispersal. Strikingly, the only two known stars with three or more planets younger than $\lesssim 100$ Myrs, HR 8799 and V1298 Tau, have been suggested to be in such MMR chains, given the orbits' near-integer period ratios.
We incorporate recent transit and radial velocity observations of the V1298 Tau system, and investigate constraints on the system's orbital architecture imposed by requiring dynamical stability on timescales much shorter than the system's age. We show that the recent radial-velocity mass measurement of V1298 Tau $b$ places it within a factor of two of the instability limit, and that this allows us to set significantly lower limits on the eccentricity ($e_b \leq 0.17$ at $99.7\%$ confidence). Additionally, we rule out a resonant chain configuration for V1298 Tau at $\gtrsim 99\%$ confidence. Thus, if the $\sim 23$ Myr-old V1298 Tau system did form as a resonant chain, it must have undergone instability and rearrangement shortly after disk dispersal. We expect that similar stability constrained characterization of future young multi-planet systems will be valuable in informing planet formation models.

\end{abstract}


\section{Introduction} \label{sec:intro}


The \textit{Kepler} mission \citep{Borucki2010} revealed a broad population of planets smaller than Neptune ($\lesssim 4\rearth$) in compact orbital configurations  \citep{Howard2012,Petigura2018}. Theoretical studies suggest that planetary systems may have originally formed with higher planetary multiplicity and short lifetimes, and undergone instabilities that rearranged them into more widely spaced and progressively longer-lived configurations \citep{Volk2015, Pu2015, Bitsch2019}. \cite{Izidoro2017, Izidoro2021} \citep[see also][]{Goldberg2022} additionally argue that dissipative migration in a protoplanetary disk should promote initial formation in chains of mean-motion resonances (MMRs) between the planetary orbits, and suggest that subsequent dynamical instabilities could scramble orbital periods, explaining the wide period-ratio distribution and dearth of chains of MMRs observed at Gyr ages \citep{Fabrycky2014, Morrison2020}. 

Targeted efforts to find young planetary systems could test this hypothesis by checking whether such systems are preferentially found in resonant chains at early ages. 
Several such searches are underway, e.g., the Zodiacal Exoplanets In time survey \citep[ZEIT;][]{Rizzuto2017} and the Cluster Difference Imaging Photometric Survey \citep[CDIPS;][]{Bouma2019}.
Although several young exoplanets have been found, e.g., AU Mic $b$ and $c$ \citep{Plavchan2020}, DS Tuc A $b$ and $c$ \citep{Newton2019}, K2-33 $b$ \citep{David2016}, there are very few young systems with three or more planets to check for resonant chains.

Intriguingly, for the only two known 3+ planet systems younger than 100 Myr, $\sim 23$ Myr old V1298 Tau \citep{David2019a, David2019b} and $\sim 40$ Myr old HR8799 \citep{Marois2008a}, it has been suggested that both may be in a resonant chain based on their observed period ratios \citep{Wang2018, Feinstein2021}.
Given that only $\sim 1 \%$ of mature systems with three or more planets are found in MMR chains \citep{Morrison2020}, such a resonant configurations for the youngest multiplanet systems would support the theoretical work of \cite{Izidoro2017}, \cite{Bitsch2019}, and \cite{Izidoro2021}.

However, whether or not these systems are in fact in MMR chains depends on not just the period ratios, but also the planetary masses and orbital eccentricities.
While radial velocity (RV) measurements are challenging due to strong activity in the young star V1298 Tau, recent RV measurements by \cite{Mascareno2021} find a mass for planet $b$ of $0.64 \pm 0.19 \mjup$, and $3\sigma$ upper limits for the masses of $c$ and $d$ of 0.24 and 0.31 $\mjup$, respectively.
The large mass measured for planet $b$ in this compact system implies that many possible orbital configurations would immediately go unstable.

We therefore investigate the constraints imposed by dynamical stability on the masses, orbital parameters, and resonant configuration of the V1298 Tau planets\footnote{Jupyter notebooks and scripts used in our analysis can be found in \url{https://github.com/Rob685/V1298Tau_stability}.}. Our work is informed by the recent RV observations from \cite{Mascareno2021}, and period ratio observations from $K2$ and \textit{TESS} transits separated by roughly 6.5 years \citep{David2019b,Feinstein2021}. 
Finally, we discuss the implications of our results for the hypothesis that planets generically form in MMR chains that subsequently destabilize \citep{Izidoro2017, Bitsch2019, Izidoro2021, Goldberg2022}. 

\section{Resonant and Near-Resonant Configurations for Planets $c$ and $d$}
\label{sec:MMRs}

The orbital periods of planets $c$, $d$, and $b$ (listed in increasing distance from the star, which differs from the order of their discovery) are precisely measured across several transits, with fractional errors  $\sim 10^{-3}$ days. 
The outer planet's (planet $e$) orbital period is significantly less certain \citep[$P_e = 50.29 \pm 6.62 $ days;][]{Feinstein2021}, since only two transits spaced by roughly 6.5 years have been observed.
We ignore planet $e$ in this study due to this uncertainty, and argue in Section~\ref{subsec:pl_e} that for reasonable planet parameters, its omission does not significantly affect our results.

We begin by sampling a wide range of masses, eccentricities and orbital orientations for the inner planets $c$ and $d$ in and near the 3:2 MMR consistent with their observed period ratio $\approx 1.503$, and in the next section explore the constraints imposed by orbital stability in the presence of planet $b$. 
Doing the analysis in this two-step process means that our full orbital configurations do not necessarily exactly match all the available orbital period measurements, but greatly clarifies which combinations of parameters are dominantly constrained by stability in the high-dimensional parameter space.
Given that \textit{Spitzer} observations (Livingston et al., \textit{in prep.}) and further RV analysis (Blunt et al. \textit{in prep.}) of the system will soon provide new constraints, we defer detailed orbit-fitting to those studies, and provide a framework for incorporating stability requirements.

\subsection{Sampling Orbits for planets c and d near the 3:2 MMR}\label{subsec:c_and_d}

Planets $c$ and $d$ transited $9$ and $5$ times during the \textit{K2} observations, and $4$ and $3$ times when observed by \textit{TESS}, roughly 6.5 years ($\approx 200$ orbits) later, respectively.
This effectively provides two `instantaneous' measurements of their physically meaningful period ratio: $1.503 \pm 0.0001$ as measured in 2015 with \textit{K2} \citep{David2019b}, and $1.503 \pm 0.0005$ as measured in 2021 with \textit{TESS} \citep{Feinstein2021}.

The simplest explanation for the consistent period ratio measurements in the two epochs is that the orbital periods are approximately constant in time, though more general oscillating solutions one might expect near MMRs \citep[See Chapter 8 of][]{Murray2000} are also possible.
Most of the techniques for speeding up Markov Chain Monte Carlo parameter estimation methods \citep[e.g.,][]{Foreman-Mackey2013} do not work for sampling the wide and disconnected sets of masses and orbital parameters consistent with two observed period ratios.
We therefore opt for a simple rejection sampling approach \citep[e.g.,][]{Price_Whelan2017} where we accept a set of masses and orbital parameters based on the likelihood $L$ of the period ratio passing through the observations\footnote{Dropping a scaling offset term that does not affect the rejection sampling.}:
\begin{equation}
    \log{L} = -\frac{1}{2}\sum_i \frac{[p_{\rm{obs}}(t_i) - p_{\rm{model}}(t_i)]^2}{\sigma_i^2},
\end{equation}
where the $t_i$ are the times of the (two) observations, $\sigma_i$ the observational uncertainties (assumed Gaussian), $p_{\rm{obs}}$ are the observed period ratios, and $p_{\rm{model}}$ are the period ratios as determined using the \texttt{WHFast} integrator \citep{Rein2015} in the \texttt{REBOUND} N-body package \citep{Rein2012}.

In order to sample a wide range of resonant and near-resonant configurations consistent with the observed period ratios, while reducing the number of parameters and aiding in their interpretation, we exploit analytical models of the planetary dynamics in and near MMRs.
In particular, we sample four parameters: the planet pair's total planet-star mass ratio $\mu = (m_c + m_d)/m_\star$, the equilibrium combined eccentricity $e_{\rm{forced}}$ parametrizing the strength of the 3:2 MMR, the combined free eccentricity $e_{\rm{free}}$, measuring how far the planet pair is from the equilibrium value, and the resonant angle $\phi$, which measures the azimuthal angle at which conjunctions between the two planets occur, and sets the phase of the period ratio oscillations.
The fact that two masses and twelve orbital parameters for the two planets can be approximately distilled into four parameters determining the period ratio evolution is not obvious, so we provide more details in the Appendix.
We perform these and all other calculations of resonant parameters in this paper using the open-source \texttt{celmech} package\footnote{Detailed API: \url{https://github.com/shadden/celmech}.}.

\subsection{Resonant and Near-resonant Configurations}\label{subsec:lnlike}

We sampled $\eforced$ and $\efree$ uniformly from 0 to 0.2 (i.e., nearly orbit-crossing), and the total planet-star mass ratio $\mu$ log-uniformly.
These planets may be significantly inflated \citep{David2019b}, so we adopt a low-mass bound of $\mu$ at 3$\mearth$ based on measurements of the least dense exoplanet known-- Kepler-51b \citep{Steffen2013, Madusa2014}. For the upper bound, we chose 1$\mjup$ to conservatively account for the RV upper bounds \citep{Mascareno2021}.

\begin{figure*}[ht!]
\centering
\includegraphics[width=1.0\textwidth]{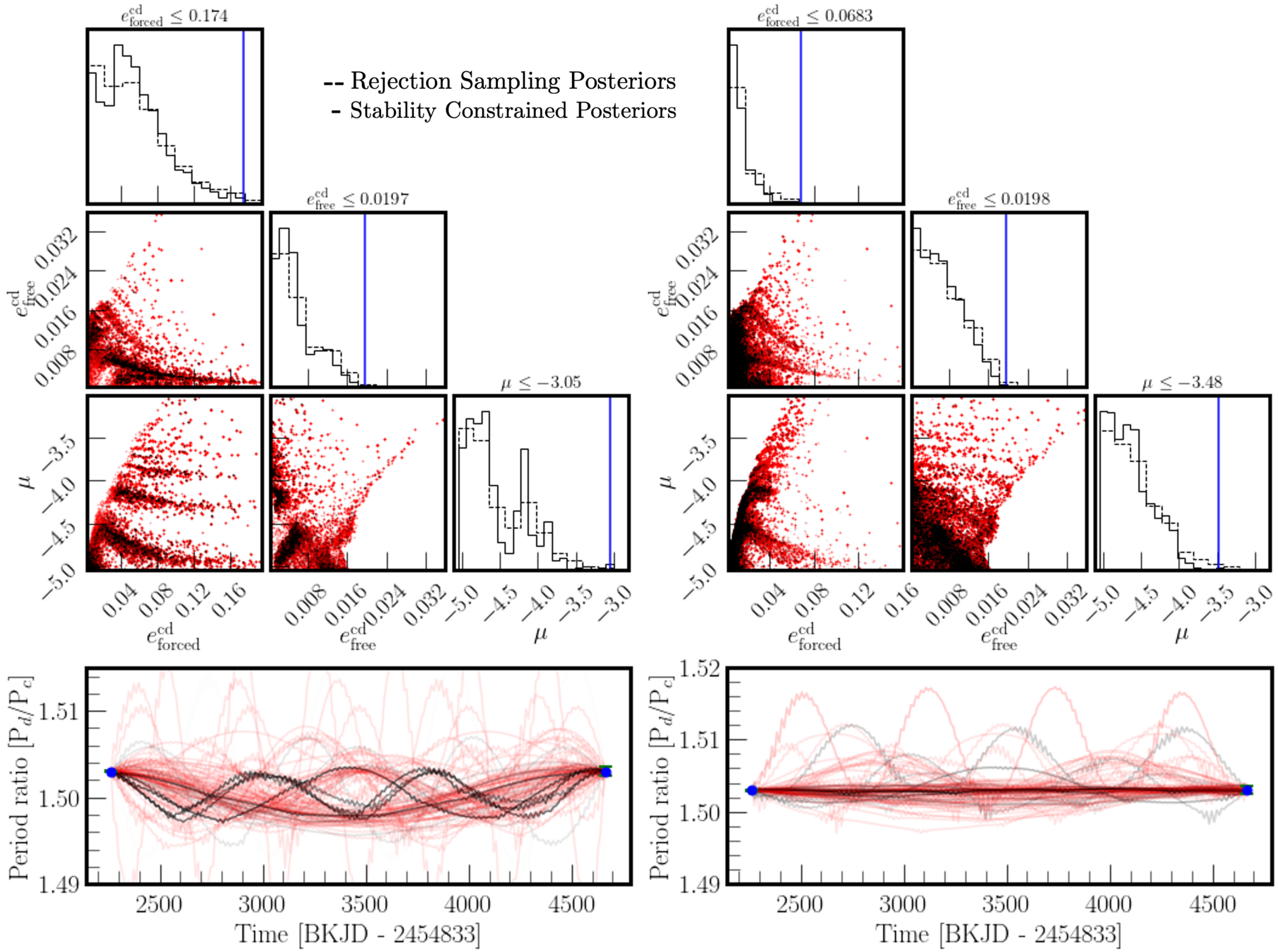}
\caption{Posterior distribution of orbital configurations for the inner two planets, $c$ and $d$, based on the two observed period ratios shown in blue in the bottom panels.
The left panel shows configurations where the planets are in the 3:2 MMR, and the right panel shows non-resonant configurations.
See the end of Section~\ref{subsec:c_and_d} for variable definitions.
Red configurations go unstable in N-body integrations within $\approx 225$ years ($10^4$ orbits) upon sampling a mass and orbit for planet $b$ from RV observations (Section~\ref{sec:stability}).
The dashed histograms trace all of the configurations, while the solid histograms are the stability constrained distrubutions, i.e.,  weighted by each configuration's probability of stability as evaluated by SPOCK (red points have SPOCK probability = 0). The blue solid lines mark the $99.7$th percentile upper limit for each parameter.
The lower panels show period ratio time series for a 100 random samples, passing through the observations, marked as blue points with their (small) error bars. 
See text for discussion.}
\label{fig:in_vs_out}
\end{figure*}

We sampled 60 million configurations, accepting 9,693. This posterior size is adequate to explore our parameter space, and is shown in Figure~\ref{fig:in_vs_out}, where the dashed histograms are the rejection posteriors. The color-coding and solid histograms incorporate stability constraints and are discussed in Section~\ref{sec:stability}. 

We divide our configurations into ones that are inside the 3:2 MMR  on the left half of Figure~\ref{fig:in_vs_out}, and ones that are not ($\approx 78\%$) on the right (see Appendix \ref{app:resvsnot} for details).
We see that in both the resonant and non-resonant cases, the observed period ratios for planets $c$ and $d$ significantly favor lower masses.
In the resonant cases, the forced eccentricities parametrizing the strength of the MMR are higher than in the non-resonance configurations.
This reflects the fact that for most of the configurations on the right of Figure~\ref{fig:in_vs_out}, the MMR is weak enough that there is no resonant region at all (Appendix \ref{app:resvsnot}).
The oscillations around the equilibrium, parametrized by $e_{\rm{free}}$ are small in both cases.

We also show the period ratio evolution for a random selection of 100 solutions in the bottom panels with the two observed values in blue.
Many of the non-resonant configurations on the right exhibit flat period ratios because the masses and orbital parameters render the MMR weak.
For the resonant cases where the MMR is strong, we get finite-amplitude oscillations in the period-ratio (bottom left panel of Figure~\ref{fig:in_vs_out}) even if we restrict $e_{\rm{free}} = 0$, which by definition should correspond to a configuration at the resonant equilibrium with a constant period ratio. 
This discrepancy is due to the fact that the analytical MMR model, and its prediction for the location of the resonant equilibrium, is only approximate.
Given that the two observations in blue (bottom panels of Figure~\ref{fig:in_vs_out}) are consistent within very narrow error bars (plotted), improved or numerical models would provide a stronger preference for nearly flat period-ratio solutions with $e_{\rm{free}}\approx0$.
Since our goal is simply to explore a representative range of resonant and near-resonant configurations for planets $c$ and $d$ in the presence of a massive planet $b$, we do not pursue this further.
A third set of transits for planets $c$ and $d$ would provide a strong and valuable constraint on their resonant configuration, and for whether the period ratio is constant with time.

For the resonant cases (left side of Figure~ \ref{fig:in_vs_out}), the configurations most vulnerable to instabilities due to perturbations from other planets are those near the boundaries of the MMR, i.e., near the separatrix \citep[e.g.,][]{Rath2021}.
The separatrix corresponds to a particular value of $e_{\rm{free}}$, or equivalently to the black `cat's-eye' boundary in the plane spanned by $\phi$ and the period ratio deviation from 3/2 shown in the left panel of Figure~ \ref{fig:catseye} (see Appendix for details). 
To account for the fact that the size of this resonant region varies with $\mu$ and $e_{\rm{forced}}$, for each of the resonant configurations, we use \texttt{celmech} to normalize the period ratio deviation by the value at the separatrix so that the MMR boundary for all the resonant configurations plotted in Figure~ \ref{fig:catseye} extends from $[-1,1]$. 
We see that while a diffuse distribution of configurations fill the `cat's eye' resonant region, most configurations cluster closer to the equilibrium (equivalently at low $e_{\rm{free}}$), with higher mass solutions closer to the equilibrium.

\begin{figure*}[!ht]
    \centering
    \includegraphics[width=1.0\textwidth]{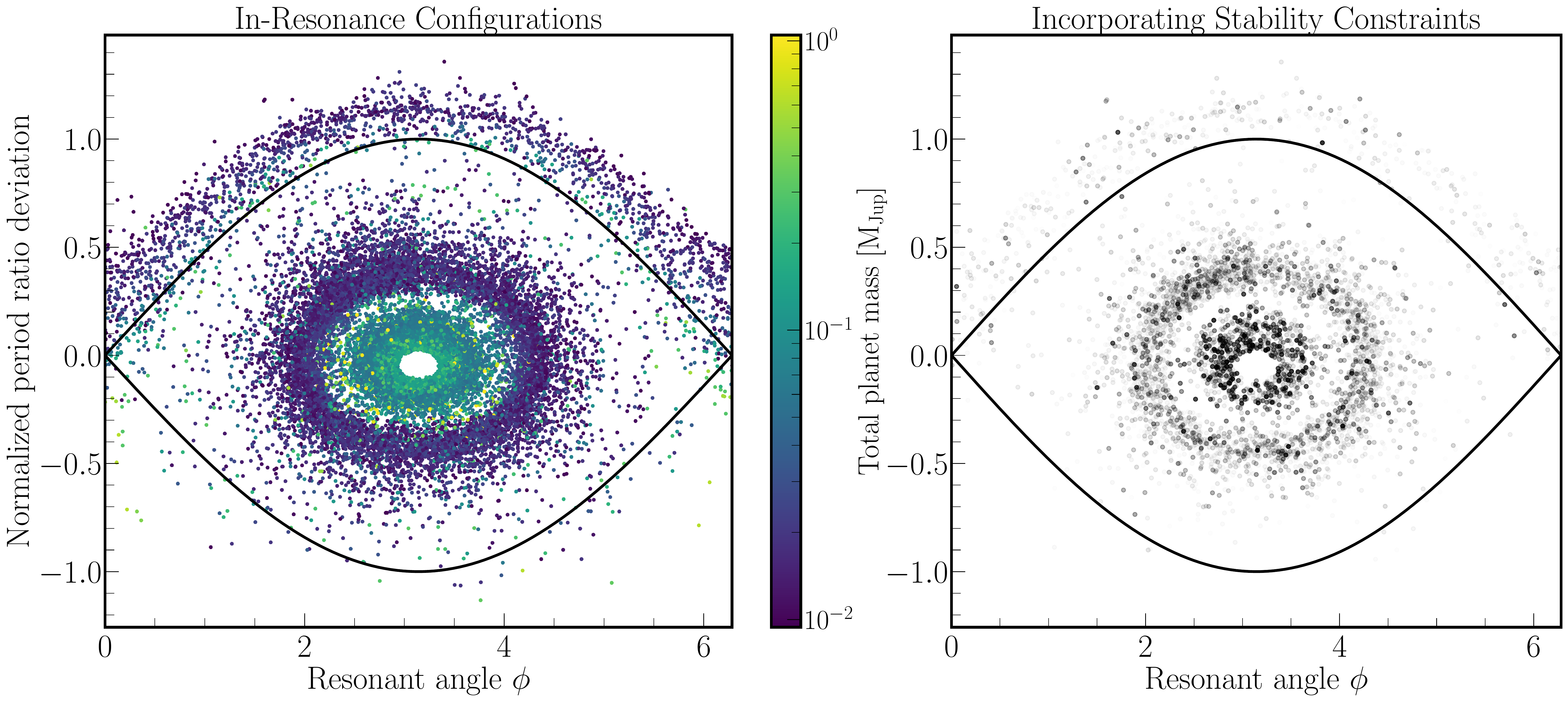}
    \caption{For the resonant configurations on the left of Figure~\ref{fig:in_vs_out}, we plot the 3:2 resonant island for planets $c$ and $d$, bounded by the `cat's eye' separatrix trajectory (solid black line).
    The $x$ axis is the 3:2 resonant angle, and the $y$ axis is the period ratio's deviation from 3/2.
    The resonance region varies for each configuration (points) since it depends on the masses and orbital parameters, so the period ratio deviation has been normalized so that the separatrix extends from $\approx [-1,1]$ for all points (see Appendix).
    Left panel shows the original posterior samples (both black and red points in Figure~\ref{fig:in_vs_out}), with the color bar denoting the total planet mass $m_c+m_d$.
    The right panel applies stability constraints by setting an opacity to each point given by the SPOCK probability of stability.
    As expected, stability preferentially removes configurations near the separatrix, but it also imposes additional structure.}
\label{fig:catseye}
\end{figure*}

\section{Stability Constrained Characterization}\label{sec:stability}

With a range of resonant and near-resonant configurations for $c$ and $d$ in hand, we now introduce planet $b$ and determine the constraints imposed by stability.
In order to make this step computationally tractable, we use the Stability of Orbital Configurations Klassifier (SPOCK) package \citep{Tamayo2020, Cranmer2021}, a collection of machine learning and analytical models for predicting the stability of compact multi-planet systems.

We randomly draw $10^5$ resonant and near-resonant configurations for planets $c$ and $d$ from the samples generated in Section \ref{subsec:lnlike} (with replacement), and introduce planet $b$. 
We note that, in general, the addition of planet $b$ causes the evolution of the period ratio between $c$ and $d$ to no longer pass exactly through the observed values of $1.503$, but we leave detailed orbit-fitting to imminent future observations (Livingston et al., in prep).
Our goal instead is to explore stability constraints for a representative sample of resonant and near-resonant configurations for planets $c$ and $d$, given the RV mass and orbital eccentricity measured by \cite{Mascareno2021}.

We choose the orbital period of planet $b$ to yield the precisely observed period ratio $P_b/P_d = 1.946$, and draw the masses and orbital eccentricities from the RV posteriors of \cite{Mascareno2021}, assuming Gaussian errors: $m_b \sim \mathcal{N}(0.64 \pm 0.19)$, $e_b \sim \mathcal{N}(0.13 \pm 0.07)$.
Finally, we sample the longitude of pericenter and true longitude of planet $b$ uniformly between 0 and $2\pi$. 
We then use SPOCK \citep{Tamayo2020} to calculate the probability of each of the $10^5$ configurations being stable for the next $\sim 10^9$ orbits.
Red configurations in Figure~\ref{fig:in_vs_out} went unstable within $10^4$ orbits (SPOCK probability of zero), and the stability constrained (solid line) histograms weight each sample by its respective SPOCK probability \citep{Tamayo2021a}.

\subsection{How long to require stability?} \label{sec:howlong}

Our approach requires closer examination for this particularly young system, given the possibility that planetary systems may generically be {\it unstable}, undergoing dynamical instabilities throughout their lives \citep[e.g.,][]{Laskar1996, Volk2015, Pu2015, Izidoro2017, Bitsch2019, Izidoro2021}.
Nevertheless, in that picture, one might expect a system observed at an arbitrary snapshot in time to always have a time to instability comparable to its present lifetime \citep{Laskar1996}.
Thus, even if planetary systems are ultimately unstable, one can still require stability on timescales much shorter than their age, since otherwise instabilities would be happening all the time.

The SPOCK \texttt{FeatureClassifier} \citep{Tamayo2020} provides a probability of stability over $10^9$ orbits of the inner-most planet, corresponding to $\approx 22$ Myr for V1298 Tau.
For this young system, that timescale is close to its age, and thus too stringent a timescale on which to require stability. 

However, most instabilities occur early, and we do not expect this to significantly bias our results. 
About $96\%$ of our tested configurations had SPOCK probabilities $\leq 0.2$.
We checked the median instability times for these low-probability configurations with SPOCK's deep learning classifier \citep{Cranmer2021}, and found
$97\%$ had predicted lifetimes $\leq 10^7$ orbits ($\lesssim 0.2$ Myr).
This preference toward short lifetimes also matches expectations from the analytic investigation of these instabilities by \cite{Tamayo2021b}.  

\subsection{Constraints on Planets c and d}\label{subsec:inner_planets}

For the non-resonant cases on the right of Figure~\ref{fig:in_vs_out}, stability sets a $99.7$th percentile upper limit for the combined mass of planets $c$ and $d$ of $\mu < 0.35\mjup$, slightly more constraining than the upper limits from the RV observations \citep{Mascareno2021}.
Stability of resonant configurations (left of Figure~\ref{fig:in_vs_out}), on the other hand, allows for values up to $0.95\mjup$ ($99.7$th percentile).
Moreover, SPOCK shows a preference toward particular values of $\mu$ (peaks in the solid histogram for $\mu$, lower right triangle plot panel on left of Figure~\ref{fig:in_vs_out}).

These modes are apparent in Figure~\ref{fig:catseye}, where in the right panel we set the opaqueness of each configuration according to its probability of stability according to SPOCK.
As expected, stability preferentially eliminates configurations near the separatrix, and selects for setups closer to the resonant equilibrium.
While we do not fully understand why stability is clearing out the annulus of configurations between the turquoise and purple rings in the left panel of Figure~\ref{fig:catseye}, we speculate that this may be due to chaos induced by secondary resonances, i.e., with the frequency of oscillation around the equilibrium point \citep[See][]{Rath2021}. 
A third measurement of the period ratio between planets $c$ and $d$ would strongly constrain the amplitude of the period-ratio oscillations, and provide a valuable constraint on these planets' resonant configuration.

\subsection{Constraints on Planet b}\label{subsec:pl_b}

\begin{figure}[ht!]
    \centering
    \includegraphics[width=0.47\textwidth]{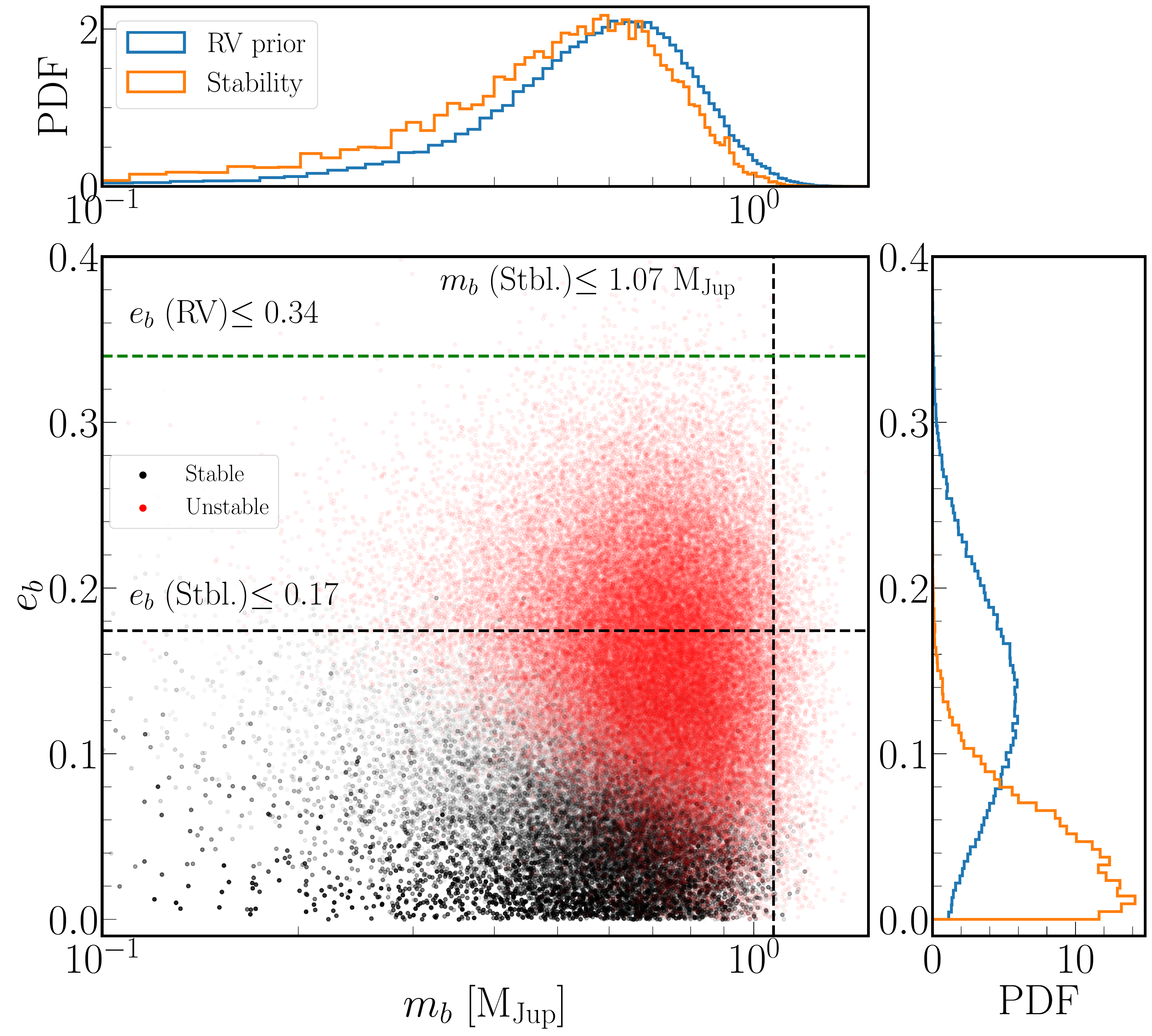}
    \caption{Stability constraints on planet $b$; masses and eccentricities were drawn from the \cite{Mascareno2021} RV posteriors (blue histograms), assuming the posteriors are Gaussian. 
    Coloring and opacity scheme of the main panel follows the same as in Figure~\ref{fig:in_vs_out}. 
    Stability (orange histograms) strongly constrains the orbital eccentricity.
    The black dashed lines represent the $99.7$th percentile upper limits from stability while the green dashed line depicts the RV Gaussian $3\sigma$ upper limit on the eccentricity (also shown in Table~\ref{tbl:posteriors}). The RV measured mass of $0.64 \pm 0.19$ is within a factor of 2 from the instability region even at zero eccentricity.}
\label{fig:eb_vs_mb}
\end{figure}

Figure~\ref{fig:eb_vs_mb} plots stability constraints on planet $b$, taking our $10^5$ samples drawn from the mass and orbital eccentricity distributions reported by \cite{Mascareno2021}. 
As with Figure~\ref{fig:in_vs_out}, red points are configurations that went unstable in N-body integrations within $10^4$ orbits, while the surviving black points have their opacity set according to their SPOCK probability of surviving for the next $\approx 20$ Myr. 

We obtain a $99.7$th percentile upper limit for $m_b$ of $1.07 \mjup$. 
Even though this upper limit is only $\sim 0.1\mjup$ lower than the Gaussian $3\sigma$ RV limit, we find that even if we extend our mass priors to higher values, masses above this limit are unstable at any eccentricity. 
Therefore, coincidentally, stability and RV observations yield very similar upper limits on planet $b$'s mass. 
We also note that the median mass estimated through RV observations of $0.64\mjup$ \citep{Mascareno2021} is within a factor of two from the stability limit even at zero eccentricity (Figure~\ref{fig:eb_vs_mb}). 

This implies strong stability constraints on planet $b$'s orbital eccentricity, $e_b$. 
We obtain a $99.7$th percentile upper limit of $0.17$ for the $e_b$, which is half of the corresponding RV limit. 
Finally, we find that the constraints on planet $b$ in Figure~\ref{fig:eb_vs_mb} do not depend strongly on the parameters for planets $c$ and $d$ (e.g., the bottom two rows of Table $\ref{tbl:posteriors}$ comparing constraints for resonant and non-resonant configurations for $c$ and $d$).
This renders our approach of examining the configuration of planets $c$ and $d$ separately from those on planet $b$ particularly valuable.

\begin{deluxetable}{cccc}
\tabletypesize{\footnotesize}
\tablecaption{$99.7$th percentile upper limits
\label{tbl:posteriors}}
\tablecolumns{4}
\tablehead{
\colhead{Parameter} &
\colhead{Inside Resonance} &
\colhead{Outside Resonance} &
\colhead{RV ($3\sigma$) reported}
}
\startdata
 $\mu$ & 0.94$\mjup$ & 0.35$\mjup$ & 0.55$\mjup$ \\ 
 $\eforced$ & 0.17 & 0.06 & - \\
 $\efree$ & 0.02 & 0.02 & - \\ 
 $m_b$ & 1.07$\mjup$ & 1.07$\mjup$ & 1.21$\mjup$ \\
 $e_b$ & 0.17 & 0.17 & 0.34
\enddata
\tablecomments{$99.7$th percentile upper limits. The second and third columns correspond to stability limits in the cases where planets $c$ and $d$ are inside and outside the 3:2 MMR, respectively (left and right of Figure~\ref{fig:in_vs_out}). The last column are the Gaussian $3\sigma$ RV limits reported by \cite{Mascareno2021}. Bootstrap re-sampling of the stability posteriors give errors on the order of $\sim 10^{-3}$ for each parameter.}
\end{deluxetable}

\subsection{Is the System in an MMR Chain?} \label{subsec:mmr_chain}
A challenge to establishing whether or not V1298 Tau is in an MMR chain is that the theory for MMR chains \citep[e.g.,][]{Delisle2017, Siegel2021} is less developed than for MMRs between a single pair of planets.
We therefore take two approximate approaches.

First, we ask how many stable configurations place planets $d$ and $b$ (observed period ratio $\approx 1.946$) inside the 2:1 MMR, assuming a two-planet model.
In particular, when counting the fraction of configurations in the 2:1 MMR, we weight each one by its corresponding SPOCK probability of stability $p_i$ \citep{Tamayo2021a}:

\begin{equation}
P(\rm{MMR\ chain}) = \sum_{i\ in\ 2:1} p_{i}/\sum_i p_i, \label{eq:pmmr}
\end{equation}
This yields $P(\rm{MMR\ chain}) \approx 1\%$.
Even if future observations yield lower masses and eccentricities for planet $b$, the 2:1 MMR region lies at higher masses, making the system unstable. The fact that $P_b/P_d \approx 1.946$, far from 2, is only in the 2:1 resonance region for the highest masses and eccentricities makes it even less likely that planets $d$ and $b$ are in the 2:1 resonance.

However, this analytical two-planet model ignores the gravitational effects of the third planet $c$, which can be important.
We therefore instead check for libration of the three-body angle $\phi_{\rm{chain}} = \lambda_c - 2\lambda_d + \lambda_b$ over timescales of $10^3$ orbits of planet $c$, where the $\lambda$ are the planets' mean longitudes. Using this libration condition to select configurations in the 2:1 MMR for the first sum of Equation~\ref{eq:pmmr}, we obtain $P(\rm{MMR\ chain}) \approx 0.02\%$.
Both tests therefore rule out an MMR chain configuration for V1298 Tau at $\gtrsim 99\%$ confidence.

\subsection{Ignoring Planet e}\label{subsec:pl_e}

Similar to the stability step performed for planet $b$, we sampled the eccentricity and mass distributions of planet $e$ from \cite{Mascareno2021} and the period distribution from \cite{Feinstein2021}. We drew $10^5$ samples from the stable configurations already with planet $b$, and we conducted uninformative and RV prior analysis similar to those we performed with planet $b$.  In both cases, we found the inclusion of planet $e$ does not affect the results of the inner planets, nor does stability constrain the period, mass, or eccentricity any better than what is already reported. The fact that plausible parameters for planet $e$ do not affect our results justifies our choice to simplify the parameter space by excluding it.

\section{Discussion and Conclusions}\label{sec:discussion}

The theoretical work of \cite{Izidoro2017, Izidoro2021} posits that dynamical instabilities could begin as soon as the proto-planetary disk dissipates at $\sim 5$ Myr for compact super-Earth systems. For systems with giant outer planets, \cite{Bitsch2019} found that dynamical instability can eject smaller bodies in the inner regions of the systems due to interactions from outer and migrating giant planets, leaving less compact but stable systems. The findings presented here provide a first suggestion that either these processes did not occur in V1298 Tau, or that the planets in this system experienced dynamical instability shortly after their natal disk vanished leaving behind near-MMR period ratios as birth marks of a violent recent past. If this framework does occur, then future discoveries of young multiplanet systems could constrain the subsequent dynamical interaction timescale.

Our conclusions are as follows:
\begin{enumerate}
    \item Stability rules out a MMR chain at $\gtrsim 99\%$ confidence, assuming the RV mass and eccentricities from \cite{Mascareno2021}. Lower values for these parameters further narrow the resonant region, diminishing the probability of planets $d$ and $b$ being in the 2:1 resonance. This implies that either the V1298 Tau planets did not form in an MMR chain, or if they did, they must have undergone a dynamical instability quickly following disk dispersal, within $\approx 20$ Myr.
    \item Figure~\ref{fig:in_vs_out} shows that the period ratio observations coupled with stability analysis can constrain the inner planet pair's combined mass. Additional sets of transits would help narrow down the resonant configuration and masses of the planets in the system.
    \item If the inner two planets are in the 3:2 MMR, higher planetary masses require the planet-pair to be closer to the resonance equilibrium (i.e., closer to the center of Figure~\ref{fig:catseye}). 
    \item The RV-measured mass for planet $b$ is within a factor of two from the instability limit (see Figure~\ref{fig:eb_vs_mb}). Stability places $99.7$th percentile upper limits of $1.1\mjup$ and $0.17$ for its mass and eccentricity, respectively.

\end{enumerate}

Future observations of V1298 Tau will provide extremely valuable information on this landmark system, and we have provided a framework for incorporating complementary constraints from stability.
While one can not make general conclusions from a single system, this case study highlights the promise of ongoing searches for young planets in constraining our understanding of how planetary systems form and evolve over their Gyr lifetimes.

\software{
          celmech \citep{Hadden2019},
          corner \citep{Foreman-Mackey2016},
          Jupyter Notebooks \citep{Kluyver2016},
          Matplotlib \citep{Hunter2007},
          NumPy \citep{numpy2011, Harris2020},
          REBOUND \citep{Rein2012}
          SciPy \citep{Virtanen2020},
          SPOCK \citep{Tamayo2020}
          }

\bibliography{references.bib}

\begin{thebibliography}{}
\expandafter\ifx\csname natexlab\endcsname\relax\def\natexlab#1{#1}\fi
\providecommand{\url}[1]{\href{#1}{#1}}
\providecommand{\dodoi}[1]{doi:~\href{http://doi.org/#1}{\nolinkurl{#1}}}
\providecommand{\doeprint}[1]{\href{http://ascl.net/#1}{\nolinkurl{http://ascl.net/#1}}}
\providecommand{\doarXiv}[1]{\href{https://arxiv.org/abs/#1}{\nolinkurl{https://arxiv.org/abs/#1}}}

\bibitem[{Bitsch {et~al.}(2019)Bitsch, Izidoro, Johansen, Raymond, Morbidelli,
  Lambrechts, \& Jacobson}]{Bitsch2019}
Bitsch, B., Izidoro, A., Johansen, A., {et~al.} 2019, A{\&}A, 623,
  \dodoi{10.1051/0004-6361/201834489}

\bibitem[{Borucki {et~al.}(2010)Borucki, Koch, Basri, \& Batalha}]{Borucki2010}
Borucki, W., Koch, D., Basri, G., \& Batalha, N. 2010, Science, 327, 977,
  \dodoi{10.1126/science.1183147}

\bibitem[{Bouma {et~al.}(2019)Bouma, Hartman, Bhatti, Winn, \&
  Bakos}]{Bouma2019}
Bouma, L.~G., Hartman, J.~D., Bhatti, W., Winn, J.~N., \& Bakos, G.~Ã. 2019,
  The Astrophysical Journal Supplement Seires, 245,
  \dodoi{10.3847/1538-4365/ab4a7e}

\bibitem[{Cranmer {et~al.}(2021)Cranmer, Tamayo, Rein, Battaglia, Hadden,
  Armitage, Ho, Spergel, Hernquist, Hadden, \& Ho}]{Cranmer2021}
Cranmer, M., Tamayo, D., Rein, H., {et~al.} 2021, PNAS, 118, 2026053118,
  \dodoi{10.1073/pnas.2026053118}

\bibitem[{David {et~al.}(2019{\natexlab{a}})David, Petigura, Luger,
  Foreman-Mackey, Livingston, Mamajek, \& Hillenbrand}]{David2019b}
David, T.~J., Petigura, E.~A., Luger, R., {et~al.} 2019{\natexlab{a}}, The
  Astrophysical Journal Letters, 885, L12, \dodoi{10.3847/2041-8213/ab4c99}

\bibitem[{David {et~al.}(2016)David, Hillenbrand, Petigura, Carpenter,
  Crossfield, Hinkley, Ciardi, Howard, Isaacson, Cody, Schlieder, Beichman, \&
  Barenfeld}]{David2016}
David, T.~J., Hillenbrand, L.~A., Petigura, E.~A., {et~al.} 2016, Nature
  Letter, 534, \dodoi{10.1038/nature18293}

\bibitem[{David {et~al.}(2019{\natexlab{b}})David, Cody, Hedges, Mamajek,
  Hillenbrand, Ciardi, Beichman, Petigura, Fulton, Isaacson, Howard,
  Gagn{\'{e}}, Saunders, Rebull, Stauffer, Vasisht, \& Hinkley}]{David2019a}
David, T.~J., Cody, A.~M., Hedges, C.~L., {et~al.} 2019{\natexlab{b}}, The
  Astronomical Journal, 158, 79, \dodoi{10.3847/1538-3881/ab290f}

\bibitem[{Deck {et~al.}(2013)Deck, Payne, \& Holman}]{Deck2013}
Deck, K.~M., Payne, M., \& Holman, M.~J. 2013, The Astrophysical Journal, 774,
  \dodoi{10.1088/0004-637X/774/2/129}

\bibitem[{Delisle(2017)}]{Delisle2017}
Delisle, J.-B. 2017, A{\&}A, 605, 96, \dodoi{10.1051/0004-6361/201730857}

\bibitem[{Fabrycky {et~al.}(2014)Fabrycky, Lissauer, Ragozzine, Rowe, Steffen,
  Agol, Barclay, Batalha, Borucki, Ciardi, Ford, Gautier, Geary, Holman,
  Jenkins, Li, Morehead, Morris, Shporer, Smith, Still, \&
  Van~Cleve}]{Fabrycky2014}
Fabrycky, D.~C., Lissauer, J.~J., Ragozzine, D., {et~al.} 2014, The
  Astrophysical Journal, 790, 146, \dodoi{10.1088/0004-637X/790/2/146}

\bibitem[{Feinstein {et~al.}(2021)Feinstein, David, Montet, Foreman-Mackey,
  Livingston, \& Mann}]{Feinstein2021}
Feinstein, A.~D., David, T.~J., Montet, B.~T., {et~al.} 2021, ApJ Letters, 925,
  L2.
\newblock \url{https://ui.adsabs.harvard.edu/abs/2021arXiv211108660F/abstract}

\bibitem[{Foreman-Mackey(2016)}]{Foreman-Mackey2016}
Foreman-Mackey, D. 2016, The Journal of Open Source Software, 1,
  \dodoi{10.21105/joss.00024}

\bibitem[{Foreman-Mackey {et~al.}(2013)Foreman-Mackey, Hogg, Lang, \&
  Goodman}]{Foreman-Mackey2013}
Foreman-Mackey, D., Hogg, D.~W., Lang, D., \& Goodman, J. 2013, Publications of
  the Astronomical Society of the Pacific, 125, 306.
\newblock \url{https://iopscience.iop.org/article/10.1086/670067/pdf}

\bibitem[{Goldberg \& Batygin(2022)}]{Goldberg2022}
Goldberg, M., \& Batygin, K. 2022, Accepted to AJ, 14pp.
\newblock \url{https://arxiv.org/pdf/2203.00801.pdf}

\bibitem[{Hadden(2019)}]{Hadden2019}
Hadden, S. 2019, The Astronomical Journal, 158, 238,
  \dodoi{10.3847/1538-3881/ab5287}

\bibitem[{Harris {et~al.}(2020)Harris, Jarrod~Millman, van~der Walt, Gommers,
  Virtanen, Cournapeau, Wieser, Taylor, Berg, Smith, Kern, Picus, Hoyer, van
  Kerkwijk, Brett, Haldane, Fern{\'{a}}ndez~del R{\'{i}}o, Wiebe, Peterson,
  G{\'{e}}rard-Marchant, Sheppard, Reddy, Weckesser, Abbasi, Gohlke, \&
  Oliphant}]{Harris2020}
Harris, C.~R., Jarrod~Millman, K., van~der Walt, S.~J., {et~al.} 2020, Nature,
  585, 357, \dodoi{10.1038/s41586-020-2649-2}

\bibitem[{Henrard(1984)}]{Henrard1984}
Henrard, J. 1984, Celestial Mechanics, 34, 255.
\newblock \url{https://link.springer.com/content/pdf/10.1007/BF01235807.pdf}

\bibitem[{Howard {et~al.}(2012)Howard, Marcy, Bryson, Jenkins, Rowe, Batalha,
  Borucki, Koch, Dunham, Gautier, Van~Cleve, Cochran, Latham, Lissauer, Torres,
  Brown, Gilliland, Buchhave, Caldwell, Christensen-Dalsgaard, Ciardi, Fressin,
  Haas, Howell, Kjeldsen, Seager, Rogers, Sasselov, Steffen, Basri,
  Charbonneau, Christiansen, Clarke, Dupree, Fabrycky, Fischer, Ford, Fortney,
  Tarter, Girouard, Holman, Johnson, Klaus, MacHalek, Moorhead, Morehead,
  Ragozzine, Tenenbaum, Twicken, Quinn, Isaacson, Shporer, Lucas, Walkowicz,
  Welsh, Boss, Devore, Gould, Smith, Morris, Prsa, Morton, Still, Thompson,
  Mullally, Endl, \& MacQueen}]{Howard2012}
Howard, A.~W., Marcy, G.~W., Bryson, S.~T., {et~al.} 2012, Astrophysical
  Journal, Supplement Series, 201, \dodoi{10.1088/0067-0049/201/2/15}

\bibitem[{Hunter(2007)}]{Hunter2007}
Hunter, J.~D. 2007, Computing in Science {\textbackslash}{\&} Engineering, 9,
  90.
\newblock
  \url{https://ieeexplore.ieee.org/stamp/stamp.jsp?tp=&arnumber=4160265}

\bibitem[{Izidoro {et~al.}(2021)Izidoro, Bitsch, Raymond, Johansen, Morbidelli,
  Lambrechts, \& Jacobson}]{Izidoro2021}
Izidoro, A., Bitsch, B., Raymond, S.~N., {et~al.} 2021, A{\&}A.
\newblock \url{https://arxiv.org/pdf/1902.08772.pdf}

\bibitem[{Izidoro {et~al.}(2017)Izidoro, Ogihara, Raymond, Morbidelli, Pierens,
  Bitsch, Cossou, \& Hersant}]{Izidoro2017}
Izidoro, A., Ogihara, M., Raymond, S.~N., {et~al.} 2017, MNRAS, 470, 1750,
  \dodoi{10.1093/mnras/stx1232}

\bibitem[{Kluyver {et~al.}(2016)Kluyver, Ragan-Kelley, P{\'{e}}rez, Granger,
  Bussonnier, Frederic, Kelley, Hamrick, Grout, Corlay, Ivanov, Avila, Abdalla,
  Willing, \& Development~Team}]{Kluyver2016}
Kluyver, T., Ragan-Kelley, B., P{\'{e}}rez, F., {et~al.} 2016, Positioning and
  Power in Academic Publishing: Players, Agents, and Agendas,
  \dodoi{10.3233/978-1-61499-649-1-87}

\bibitem[{Laskar(1996)}]{Laskar1996}
Laskar, J. 1996, Celestial Mechanics and Dynamical Astronomy, 64, 115.
\newblock \url{https://link.springer.com/content/pdf/10.1007/BF00051610.pdf}

\bibitem[{Lichtenberg \& Lieberman(1992)}]{Lichtenberg1992}
Lichtenberg, A., \& Lieberman, M. 1992, {Regular and Chaotic Dynamics}, 2nd
  edn., ed. F.~John, J.~Mardsen, \& L.~Sirovich (Springer Science +Business
  Media, LLC).
\newblock \url{https://link.springer.com/book/10.1007/978-1-4757-2184-3}

\bibitem[{Marois {et~al.}(2008)Marois, Macintosh, Barman, Zuckerman, Song,
  Patience, Lafren{\`{i}}, \& Doyon}]{Marois2008a}
Marois, C., Macintosh, B., Barman, T., {et~al.} 2008, Science, 322.
\newblock \url{https://www.science.org/doi/10.1126/science.1166585}

\bibitem[{Masuda(2014)}]{Madusa2014}
Masuda, K. 2014, The Astrophysical Journal, 783, 53,
  \dodoi{10.1088/0004-637X/783/1/53}

\bibitem[{Morrison {et~al.}(2020)Morrison, Dawson, \& Macdonald}]{Morrison2020}
Morrison, S.~J., Dawson, R.~I., \& Macdonald, M. 2020, ApJ, 904, 157,
  \dodoi{10.3847/1538-4357/abbee8}

\bibitem[{Murray \& Dermott(1999)}]{Murray2000}
Murray, C.~D., \& Dermott, S.~F. 1999, {Solar System Dynamics} (Cambridge
  University Press), \dodoi{10.1017/cbo9781139174817}

\bibitem[{Newton {et~al.}(2019)Newton, Mann, Tofflemire, Pearce, Rizzuto,
  Vanderburg, Martinez, Wang, Ruffio, Kraus, Johnson, Thao, Wood, Rampalli,
  Nielsen, Collins, Dragomir, Anderson, Barclay, Brown, Feiden, Hart, Isopi,
  Kielkopf, Mallia, Nelson, Rodriguez, Stockdale, Waite, Wright, Lissauer,
  Ricker, Vanderspek, Latham, Seager, Winn, Jenkins, Bouma, Burke, Davies,
  Fausnaugh, Li, Morris, Mukai, Villase{\~{n}}or, Villeneuva, De~Rosa,
  Macintosh, Mengel, Okumura, \& Wittenmyer}]{Newton2019}
Newton, E.~R., Mann, A.~W., Tofflemire, B.~M., {et~al.} 2019, The Astrophysical
  Journal Letters, 880, 17, \dodoi{10.3847/2041-8213/ab2988}

\bibitem[{Petigura {et~al.}(2018)Petigura, Marcy, Winn, Weiss, Fulton, Howard,
  Sinukoff, Isaacson, Morton, \& Johnson}]{Petigura2018}
Petigura, E.~A., Marcy, G.~W., Winn, J.~N., {et~al.} 2018, The Astronomical
  Journal, 155, 89, \dodoi{10.3847/1538-3881/aaa54c}

\bibitem[{Plavchan {et~al.}(2020)Plavchan, Barclay, Gagn{\'{e}}, Gao, Cale,
  Matzko, Dragomir, Quinn, Feliz, Stassun, M~Crossfield, Berardo, Latham, Tieu,
  Anglada-Escud{\'{e}}, Ricker, Vanderspek, Seager, Winn, Jenkins, Rinehart,
  Krishnamurthy, Dynes, Doty, Adams, Afanasev, Beichman, Bottom, Bowler,
  Brinkworth, Brown, Cancino, Ciardi, Clampin, Clark, Collins, Davison,
  Foreman-Mackey, Furlan, Gaidos, Geneser, Giddens, Gilbert, Hall, Hellier,
  Henry, Horner, Howard, Huang, Huber, Kane, Kenworthy, Kielkopf, Kipping,
  Klenke, Kruse, Latouf, Lowrance, Mennesson, Mengel, Schlieder, Tanner, Teske,
  Tinney, Vanderburg, von Braun, Walp, Wang, Xuesong~Wang, Weigand, White,
  Wittenmyer, Wright, Youngblood, Zhang, \& Zilberman}]{Plavchan2020}
Plavchan, P., Barclay, T., Gagn{\'{e}}, J., {et~al.} 2020, Nature, 582, 45,
  \dodoi{10.1038/s41586-020-2400-z}

\bibitem[{Price-Whelan {et~al.}(2017)Price-Whelan, Hogg, Foreman-Mackey, \&
  Rix}]{Price_Whelan2017}
Price-Whelan, A.~M., Hogg, D.~W., Foreman-Mackey, D., \& Rix, H.-W. 2017, ApJ,
  837, 20.
\newblock \url{https://iopscience.iop.org/article/10.3847/1538-4357/aa5e50/pdf}

\bibitem[{Pu \& Wu(2015)}]{Pu2015}
Pu, B., \& Wu, Y. 2015, The Astrophysical Journal, 807, 44,
  \dodoi{10.1088/0004-637X/807/1/44}

\bibitem[{Rath {et~al.}(2021)Rath, Hadden, \& Lithwick}]{Rath2021}
Rath, J., Hadden, S., \& Lithwick, Y. 2021

\bibitem[{Rein \& Liu(2012)}]{Rein2012}
Rein, H., \& Liu, S.-F. 2012, A{\&}A, 537, 128,
  \dodoi{10.1051/0004-6361/201118085}

\bibitem[{Rein \& Tamayo(2015)}]{Rein2015}
Rein, H., \& Tamayo, D. 2015, MNRAS, 452, 376, \dodoi{10.1093/mnras/stv1257}

\bibitem[{Rizzuto {et~al.}(2017)Rizzuto, Mann, Vanderburg, Kraus, \&
  Covey}]{Rizzuto2017}
Rizzuto, A.~C., Mann, A.~W., Vanderburg, A., Kraus, A.~L., \& Covey, K.~R.
  2017, The Astronomical Journal, 154, \dodoi{10.3847/1538-3881/aa9070}

\bibitem[{Sessin(1983)}]{Sessin1983}
Sessin, W. 1983, Celestial Mechanics, 31, 109.
\newblock \url{https://link.springer.com/content/pdf/10.1007/BF01686812.pdf}

\bibitem[{Siegel \& Fabrycky(2021)}]{Siegel2021}
Siegel, J.~C., \& Fabrycky, D. 2021, The Astronomical Journal, 161,
  \dodoi{10.3847/1538-3881/abf8a6}

\bibitem[{Steffen {et~al.}(2013)Steffen, Fabrycky, Agol, Ford, Morehead,
  Cochran, Lissauer, Adams, Borucki, Bryson, Caldwell, Dupree, Jenkins,
  Robertson, Rowe, Seader, Thompson, \& Twicken}]{Steffen2013}
Steffen, J.~H., Fabrycky, D.~C., Agol, E., {et~al.} 2013, MNRAS, 428, 1077,
  \dodoi{10.1093/mnras/sts090}

\bibitem[{Su{\'{a}}rez~Mascare{\~{n}}o
  {et~al.}(2021)Su{\'{a}}rez~Mascare{\~{n}}o, Damasso, Lodieu, Sozzetti,
  S~B{\'{e}}jar, Benatti, Zapatero~Osorio, Micela, Rebolo, Desidera, Murgas,
  Claudi, Gonz{\'{a}}lez~Hern{\'{a}}ndez, Malavolta, del Burgo, Amado, Locci,
  Tabernero, Marzari, Aguado, Turrini, Cardona~Guill{\'{e}}n,
  Toledo-Padr{\'{o}}n, Maggio, Aceituno, Bauer, Caballero, Chinchilla,
  Esparza-Borges, Gonz{\'{a}}lez-{\'{A}}lvarez, Granzer, Luque, Mart{\'{i}}n,
  Nowak, Oshagh, Pall{\'{e}}, Parviainen, Quirrenbach, Reiners, Ribas,
  Strassmeier, Weber, \& Mallonn}]{Mascareno2021}
Su{\'{a}}rez~Mascare{\~{n}}o, A., Damasso, M., Lodieu, N., {et~al.} 2021,
  Nature Astronomy Letters, \dodoi{10.1038/s41550-021-01533-7}

\bibitem[{Tamayo {et~al.}(2021{\natexlab{a}})Tamayo, Gilbertson, \&
  Foreman-Mackey}]{Tamayo2021a}
Tamayo, D., Gilbertson, C., \& Foreman-Mackey, D. 2021{\natexlab{a}}, Monthly
  Notices of the Royal Astronomical Society, 501, 4798,
  \dodoi{10.1093/mnras/staa3887}

\bibitem[{Tamayo {et~al.}(2021{\natexlab{b}})Tamayo, Murray, Tremaine, \&
  Winn}]{Tamayo2021b}
Tamayo, D., Murray, N., Tremaine, S., \& Winn, J. 2021{\natexlab{b}}, The
  Astronomical Journal, 162, 220, \dodoi{10.3847/1538-3881/ac1c6a}

\bibitem[{Tamayo {et~al.}(2020)Tamayo, Cranmer, Hadden, Rein, Battaglia,
  Obertas, Armitage, Ho, Spergel, Gilbertson, Hussain, Silburt, Jontof-Hutter,
  \& Menou}]{Tamayo2020}
Tamayo, D., Cranmer, M., Hadden, S., {et~al.} 2020, Proceedings of the National
  Academy of Sciences of the United States of America, 117, 18194,
  \dodoi{10.1073/pnas.2001258117}

\bibitem[{Van Der~Walt {et~al.}(2011)Van Der~Walt, Colbert, \&
  Varoquaux}]{numpy2011}
Van Der~Walt, S., Colbert, S.~C., \& Varoquaux, G. 2011, Computing in Science
  and Engineering, 13, 22, \dodoi{10.1109/MCSE.2011.37}

\bibitem[{Virtanen {et~al.}(2020)Virtanen, Gommers, Oliphant, Haberland, Reddy,
  Cournapeau, Burovski, Peterson, Weckesser, Bright, van~der Walt, Brett,
  Wilson, Jarrod~Millman, Mayorov, J~Nelson, Jones, Kern, Larson, Carey, Polat,
  Feng, Moore, VanderPlas, Laxalde, Perktold, Cimrman, Henriksen, Quintero,
  Harris, Archibald, Ribeiro, Pedregosa, \& van Mulbregt}]{Virtanen2020}
Virtanen, P., Gommers, R., Oliphant, T.~E., {et~al.} 2020, Nature Methods,
  \dodoi{10.1038/s41592-019-0686-2}

\bibitem[{Volk \& Gladman(2015)}]{Volk2015}
Volk, K., \& Gladman, B. 2015, The Astrophysical Journal Letters, 806, L26,
  \dodoi{10.1088/2041-8205/806/2/L26}

\bibitem[{Wang {et~al.}(2018)Wang, Graham, Dawson, Fabrycky, De~Rosa, Pueyo,
  Konopacky, Macintosh, Marois, Chiang, Ammons, Arriaga, Bailey, Barman,
  Bulger, Chilcote, Cotten, Doyon, Duch{\^{e}}ne, Esposito, Fitzgerald,
  Follette, Gerard, Goodsell, Greenbaum, Hibon, Hung, Ingraham, Kalas, Larkin,
  Maire, Marchis, Marley, Metchev, Millar-Blanchaer, Nielsen, Oppenheimer,
  Palmer, Patience, Perrin, Poyneer, Rajan, Rameau, Rantakyr{\"{o}}, Ruffio,
  Savransky, Schneider, Sivaramakrishnan, Song, Soummer, Thomas, Wallace,
  Ward-Duong, Wiktorowicz, \& Wolff}]{Wang2018}
Wang, J.~J., Graham, J.~R., Dawson, R., {et~al.} 2018, The Astronomical
  Journal, 156, 192, \dodoi{10.3847/1538-3881/aae150}

\end{thebibliography}

\appendix
\section{MMR model}
\subsection{Reducing the Dimensionality}

The dynamics near first-order MMRs depend on the orbital period ratio, masses $m_i$, and eccentricity vectors ${\bf e}_i$ (pointing in the direction toward the pericenter with a magnitude given by the eccentricity), where the subscript $i$ indexes the planet.
Two approximations simplify how the period ratio evolution depends on system parameters.
First, \cite{Deck2013} show that the period ratio dynamics depend approximately only on the total mass of the planet pair divided by the stellar mass, $\mu = (m_1 + m_2)/m_\star$ (and not the mass ratio $m_1/m_2$).
For concreteness, we set planets $c$ and $d$ to have equal masses, but we have checked that varying the mass ratio does not significantly affect our results.

Second, the period ratio evolution depends primarily on a single linear combination of the eccentricity vectors $\bm{e_-}$ \citep{Sessin1983, Henrard1984, Hadden2019},

\begin{equation}
\bm{e_-} = \sqrt{2} \: \frac{f z_1+g z_2}{\sqrt[]{f^2+g^2}} \approx {\bf e_2} - {\bf e_1}, \label{Z}
\end{equation}
where $f$ and $g$ are Fourier coefficients in the disturbing function expansion of the interplanetary potential \citep[e.g.,][]{Hadden2019}.
The approximation that $\bm{e_-}$ is nearly the vector difference ${\bf e_2} - {\bf e_1}$ is excellent for all first-order MMRs {\it except} the 2:1 (in which the relationship between $f$ and $g$ is altered by additional indirect terms that need to be considered), and can thus be thought of as a relative or anti-aligned eccentricity.
This transformation additionally defines a combined pericenter $\varpi_-$, which approximately specifies the longitude at which the two orbits come closest to one another. 

The remaining linear combination of eccentricities, which can be expressed as an approximately conserved center-of-mass eccentricity \citep[e.g.,][]{Hadden2019}, to excellent approximation does not affect the period ratio evolution for a single pair of planets on short timescales.
However, larger values of this center-of-mass eccentricity do affect stability in multi-planet configurations, by leading to larger secular oscillations in $e_-$ on longer timescales \citep{Tamayo2021b}.
In order to limit the parameter space, we set the center-of-mass eccentricity to zero as a best-case scenario for stability.
Our derived upper limits on masses and orbital parameters in this best case are thus conservative and reliable.

\subsection{Physically Meaningful Parameters}

For a  period ratio near an MMR commensurability, different masses, eccentricities, and pericenter orientations will lead to oscillations in the period.
However, there always exists a resonant equilibrium corresponding to a particular combination of the above parameters, where the period ratio remains constant.

To a good approximation, for a given $j:j-1$ MMR, the equilibrium, or forced, eccentricity $e_{\rm{forced}}$ is only a function of the equilibrium period ratio and $\mu$ \citep[e.g.,][]{Hadden2019}.
In addition the equilibrium resonant angle $\phi = j\lambda_2 - (j-1)\lambda_1 - \varpi_-$, which approximately measures the location at which conjunctions occur, is always at $\phi = \pi$. 
This corresponds to conjunctions occurring at the location where the two orbits are farthest from one another.

Initial conditions near, but not at, the resonant equilibrium will oscillate around it.
The anti-aligned eccentricity $e_-$ can thus be profitably decomposed into free and forced components  \citep[e.g.,][]{Murray2000},
\begin{equation}
    e_- = e_{\rm{forced}} + e_{\rm{free}}
\label{eq:zfree}
\end{equation}

In this picture, $e_{\rm{forced}}$ is the equilibrium value at which the anti-aligned eccentricity and period ratio would remain constant (if $\phi=\pi$), and the free eccentricity $e_{\rm{free}}$ is how far away the eccentricity is from the fixed point. 
Additionally, the forced eccentricity is a conserved quantity involving both $e_-$ and the period ratio's deviation from the resonant value (in this case $P_d/P_c = 3/2$), which lets one switch back and forth between the two variables.

Because various configurations with different resonance strengths can have very different values of $e_{\rm{forced}}$, but the equilibrium period ratio is always approximately 3/2, we choose in Figure~ \ref{fig:catseye} to plot the period ratio deviation $\delta P$.
The center at $(\phi, \delta P) \approx (\pi, 0)$ is the equilibrium, corresponding to different non-zero values of $e_{\rm{forced}}$ for each configuration (point). 
Points near the equilibrium approximately execute circles around the fixed point, allowing for a simple scalar distance from the fixed point to parametrize the oscillation amplitude. At larger distances from the resonant fixed point, the trajectories become more deformed, and can even oscillate around a different fixed point.
Since the distance from the equilibrium therefore varies for general, non-circular trajectories, we define $e_{\rm{free}}$ to be the value for the trajectory (through Equation~~\ref{eq:zfree}) when the resonant angle $\phi$ crosses through $\pi$, and we convert it to a period deviation with \texttt{celmech} to make Figure~\ref{fig:catseye}.

In order to account for the phase of oscillations to match the observed period ratios (bottom panels of Figure~ \ref{fig:in_vs_out}), we initialize the system with total mass ratio $\mu$, and $e_{\rm{forced}}$ and $e_{\rm{free}}$ at $\phi=\pi$ (i.e., along a vertical line in Figure~~\ref{fig:catseye}). We then integrate the orbits using the REBOUND N-body integrator \citep{Rein2012} for a uniformly drawn time $\Delta T$, which we marginalize over when presenting the posterior distributions in Figure~ \ref{fig:in_vs_out}.

\subsection{Resonant vs Non-Resonant Configurations} \label{app:resvsnot}

While libration of the resonant angle $\phi$ is often used a test of whether a pair of planets is in resonance, this distinction is not physically meaningful when the MMR is weak (i.e., in these cases there is no dynamically meaningful difference in behavior between a configuration where $\phi$ circulates, and one where it librates with large amplitude).
A more meaningful distinction is to separate this question into two parts.

First, there is a threshold strength for first-order MMRs (parametrized by $\mu$ and $e_{\rm{forced}}$), beyond which a separatrix appears \citep[e.g.,][]{Henrard1984, Deck2013}.
This separatrix trajectory is (in the analytic approximation of the MMR model) an infinite-period trajectory (black `cat's-eye' curve in Figure~~\ref{fig:catseye}).
Configurations inside the separatrix (small $e_{\rm{free}}$) are resonant, and cleanly separated from non-resonant trajectories (large $e_{\rm{free}}$) on the outside.
This is a physically important boundary because trajectories near the separatrix are most susceptible to chaos under perturbation \citep[e.g.,][]{Lichtenberg1992, Rath2021}.
When the MMR is weak, i.e., below the threshold MMR strength, there is no separatrix, and there is no dynamically meaningful boundary between resonant and non-resonant configurations.

We therefore use the \texttt{celmech} package to calculate whether a separatrix exists or not for each of the orbital configurations in Figure~\ref{fig:in_vs_out}.
If it does, we use the fact that the numerical value of the (conserved) Hamiltonian for a resonant trajectory (inside the separatrix) is always smaller than the value on the separatrix.
On the left of Figure~\ref{fig:in_vs_out}, we then plot all the configurations where a) a separatrix exists and b) the value of the Hamiltonian places it inside the resonant region.
All remaining configurations are then plotted on the right of Figure~\ref{fig:in_vs_out}, i.e., 1) cases where the MMR is so weak there is no separatrix and 2) cases where a separatrix does exist, but the system is outside the resonant region. 
\end{document}